\documentclass[prb,aps,twocolumn,showpacs,amsmath,amssymb]{revtex4}
\usepackage{graphicx}
\usepackage{appendix}
\usepackage{amsmath}

\begin{document}

\title{Magnetic anisotropy of FePt: effect of lattice distortion and chemical disorder}

\author{C.J.~Aas$^1$}
\author{L.~Szunyogh$^2$}
\author{J.S.~Chen$^3$ }
\author{R.W.~Chantrell$^1$}
\affiliation{$^1$ Department of Physics, University of York, York YO10 5DD, United Kingdom}
\affiliation{$^2$ Department of Theoretical Physics, Budapest University of Technology and Economics, Budafoki \'ut 8. H1111 Budapest, Hungary}
\affiliation{$^3$ Department of Materials Science and Engineering, National University of Singapore, 117576, Singapore}

\date{\today}

\begin{abstract}
We perform first principles calculations of the magnetocrystalline anisotropy energy in the ﬁve L1$_0$ FePt samples studied experimentally by Ding et al. [J. App. Phys. 97, 10H303 (2005)]. The effect of temperature-induced spin fluctuations is estimated by scaling the MAE down according to previous Langevin dynamics simulations. Including chemical disorder as given in experiment, the experimental correlation between MAE and lattice mismatch is qualitatively well reproduced. Moreover we determine the chemical order parameters that reproduce exactly the experimental MAE of each sample. We conclude that the MAE is determined by the chemical disorder rather than by lattice distortion.
\end{abstract}


\maketitle

Due to its extraordinarily high magnetocrystalline anisotropy energy (MAE), L1$_0$ FePt is of considerable interest to the development of ultrahigh density magnetic recording applications and spintronics devices. From a theoretical point of view, there is an obvious need for a complete first principles model of FePt to be used in generating effective spin Hamiltonians for the purpose of atomistic and multiscale modelling. Amongst many other issues, this requires an understanding of the role of interfacial effects and chemical disorder. The large effect of chemical disorder on the MAE of FePt has already been outlined both experimentally\cite{disorder1} and theoretically\cite{disorder2}.

Recently, the experiments were extended to thin films of FePt deposited on different substrates\cite{expl}.  A strong correlation was revealed between the MAE of the FePt sample and the lattice mismatch of the FePt films with respect to the substrate\cite{expl}. The experimental data are summarised in Table \ref{table:samples}. The chemical order parameter, $s$, is defined as the probability of finding an Fe atom on a nominal Fe site or, equivalently, as the probability of finding a Pt atom on a nominal Pt site. In the experiment, the chemical order parameters were derived from the X-ray diffraction intensities $I(001)$ and $I(002)$, ($(xyz)$ denoting the plane of diffraction), through the relationship\cite{sqrt-l10,disorder1} $s \sim \sqrt{I(001)/I(002)} \:$ and normalizing $s$ to unity for sample no.~3. We refer to the experimentally obtained chemical order parameters as $s_e$ for distinction from the chemical order parameters $s$ obtained later by fitting calculated MAE-values to experiment.

\begin{table}
\caption{Summary of experimental results by Ding et al.\cite{expl}; lattice parameters $a$ and $c$,
chemical order parameter, $s$, magnetocrystalline anisotropy energy per formula unit, $K$,
and diffraction intensity ratio, $I(001)/I(002)$.}
\begin{center}
\begin{tabular}{| c | c | c | c | c | c |}
 \hline
Sample & a (\AA) & c (\AA) & $s_e$  & $K$ (meV)  & $I(001)/I(002)$\\
 \hline
 1 & 3.88673 & 3.69977 & 0.709 & 0.493 & 1\\
 \hline
2 & 3.88279 & 3.69387 & 0.978 &0.696& 1.9\\
 \hline
3 & 3.89752 & 3.68964 & 1.000 & 0.841& 1.985\\
 \hline
4 & 3.89646 & 3.69175 & 0.965 & 0.788&1.85\\
 \hline
5 & 3.86954 & 3.71378 & 0.615 &0.271&0.7536\\
 \hline
\end{tabular}
\label{table:samples}
\end{center}
\end{table}
The aim of the present work is to investigate in detail the effect of lattice distortion and chemical order on the MAE of FePt by means of the relativistic Korringa-Kohn-Rostoker\cite{Korringa,rostoker,KKR1} method as combined with the coherent potential approximation\cite{soven,kkrcpa} (KKR-CPA). In order to differentiate between the two main properties characterizing the samples, namely, the lattice distortion and the chemical disorder, we perform calculations with and without the inclusion of chemical disorder. We then fit the calculated MAE to the experimental values using the chemical order parameter, $s$, as a fitting parameter and draw conclusions from the results of our calculations. We find that chemical disorder of each sample is the more important factor in determining the experimental\cite{expl} MAE.

%\section{Theory and computational details}
As the relativistic KKR method is well documented in the literature (see e.g.~\onlinecite{KKR1}), here we merely describe some details of our calculations. We used Density Functional Theory within the Local Spin-Density Approximation (LSDA) as parametrised by Vosko \emph {et al.}\cite{voskoCJP80}. The effective potentials and fields were treated within the atomic sphere approximation (ASA). As the thin-film samples in the experiment had a thickness of approximately 20 nm (60 formula units),\cite{expl} surface contributions to the MAE should be negligibly small. We therefore modelled the FePt samples as face-centered-tetragonal (fct) bulk lattices with lattice constants as displayed in Table  \ref{table:samples}. The self-consistent calculations were performed by using the scalar-relativistic approximation, i.e., by neglecting spin-orbit coupling\cite{SOCscaling} and solving the Kohn-Sham-Dirac equation using a spherical wave expansion up to an angular momentum quantum number of $\ell=3$. As in earlier theoretical work,\cite{disorder2} we used the coherent potential approximation (CPA) to elucidate long-range chemical disorder effects in FePt. In combination with KKR, the CPA has proved particularly useful in calculating the physical properties of chemically disordered alloys\cite{kkrcpa}. The partially disordered FePt alloy is modelled by a stack of alternating layers with the chemical compositions of Fe$_s$Pt$_{1-s}$ and Pt$_s$Fe$_{1-s}$.

The MAE is then evaluated using the \emph{magnetic force theorem}\cite{Jansen99}, which states that the difference in a system's total energy for two different directions of magnetization can be approximated by the corresponding difference of the band energies, neglecting further self-consistency, i.e.,~keeping the effective potentials and fields fixed. From previous experience we know that for transition metal systems these potentials and fields can safely be taken from self-consistent scalar-relativistic calculations\cite{KKR1}. In order to achieve a relative accuracy within 5 \% for the MAE, the associated energy integration was performed by sampling 20 energy points along a
semi-circular contour in the upper complex half-plane.  At the energy point closest to the real axis the $k$-integration was calculated using 5050 $k$-points in the irreducible segment of the two-dimensional Brillouin zone.

As the MAE should vanish at the Curie temperature, it is a rapidly decreasing function of temperature. Whilst the temperature dependence of the MAE of ordered FePt has been previously calculated in terms of different theoretical methods\cite{mryasov,MAE-T-Staunton}, in the present work we do not make an attempt to carry out a similar process, since site-resolved information is currently not available for a chemically disordered system. Instead, for an approximate comparison with experiments at room temperature, we use the scaling obtained for perfectly ordered L1$_0$ FePt in terms of Langevin dynamics simulations\cite{mryasov}, namely, $K_{T=293K} \sim 0.6 \, K_{T=0K}$.

\begin{figure}[ht]
\begin{center}
\includegraphics[scale=0.32, trim = 24mm 15mm 12mm 20mm, clip]{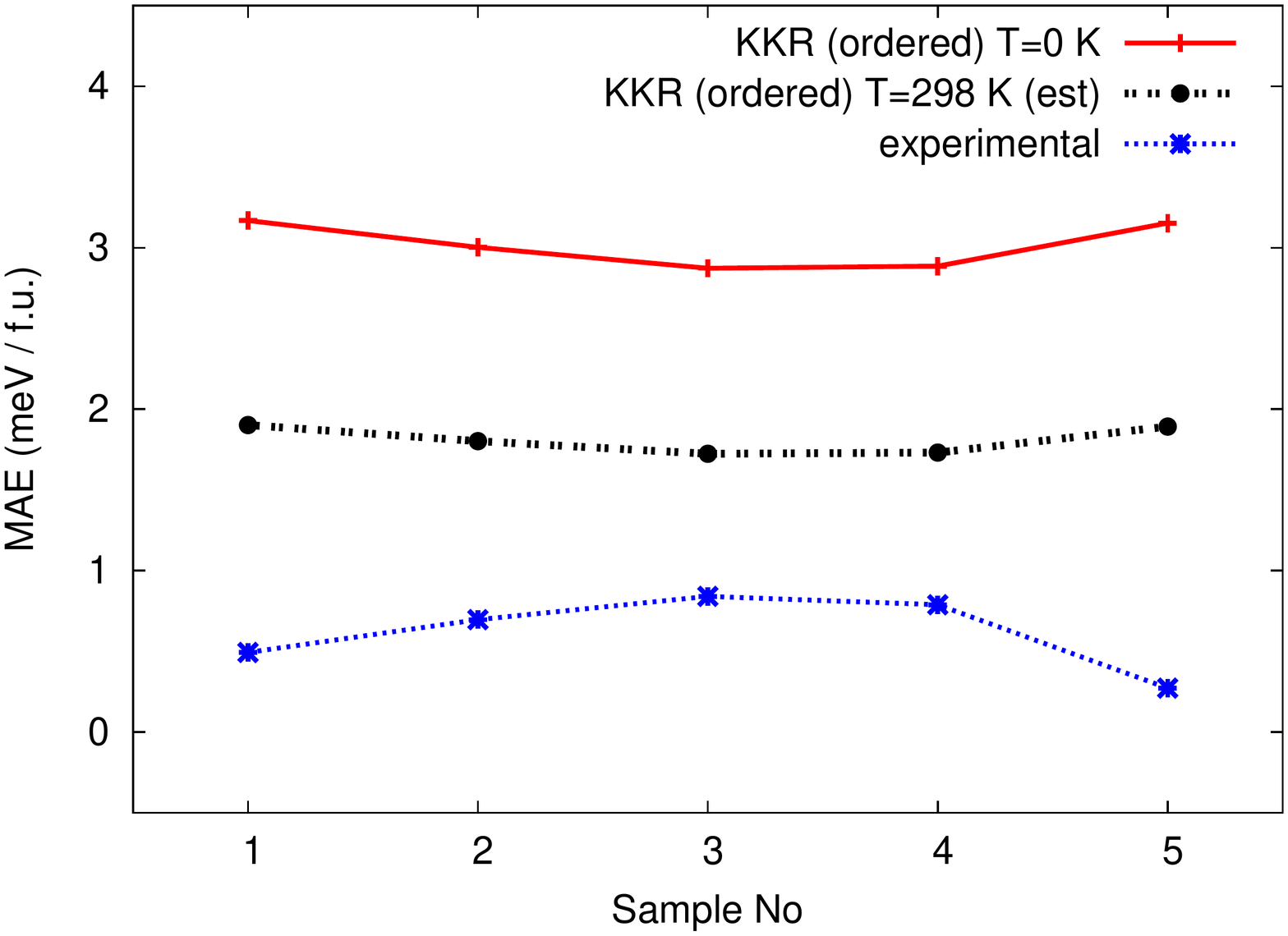}
\end{center}
\vskip -8pt
\caption{Crosses (solid line): Calculated MAE per formula unit for each of the FePt samples
in Table \ref{table:samples} modelled as perfectly ordered alloys. Circles (dashed line):
The same values scaled down by a factor of 0.6 in order to account for temperature induced effects.
Stars (dotted line): The experimental values. \label{MAE_ordered} }
\end{figure}

%\section{Results}
Using the methods described above we performed systematic calculations of the MAE of each of the FePt samples in Table \ref{table:samples}. In order to separate the effects of the lattice distortion and the chemical disorder, we split our study into three stages. In our first set of calculations, the FePt samples were modelled as perfectly ordered alloys with lattice parameters according to Table \ref{table:samples}.  As can be inferred from Fig.~\ref{MAE_ordered}, our calculated values spread around 3 meV/f.u. and show a very minor dependence on the variation of the lattice parameters.  Moreover, this moderate variation between the samples is contrary to the experimentally observed trend.

Although high in comparison to experiment, our calculated MAE values are in good agreement with other theoretical results based on the LSDA or the LSDA+U approach\cite{shick}. One obvious reason for the discrepancy between the
theoretical and experimental values is the strong temperature dependence of the MAE. We estimate this contribution by scaling the calculated MAE down by an approximate factor of 0.6, as described above. The corresponding MAE-values (also shown in Fig.~\ref{MAE_ordered}) are still too high as compared to experiment. Thus we conclude that, even when taking temperature-induced spin fluctuations into account, lattice distortion alone can explain neither the size nor the trend of the MAE obtained in the experiment.

\begin{figure}[htb]
\begin{center}
\includegraphics[scale=0.32, trim = 24mm 15mm 12mm 20mm, clip]{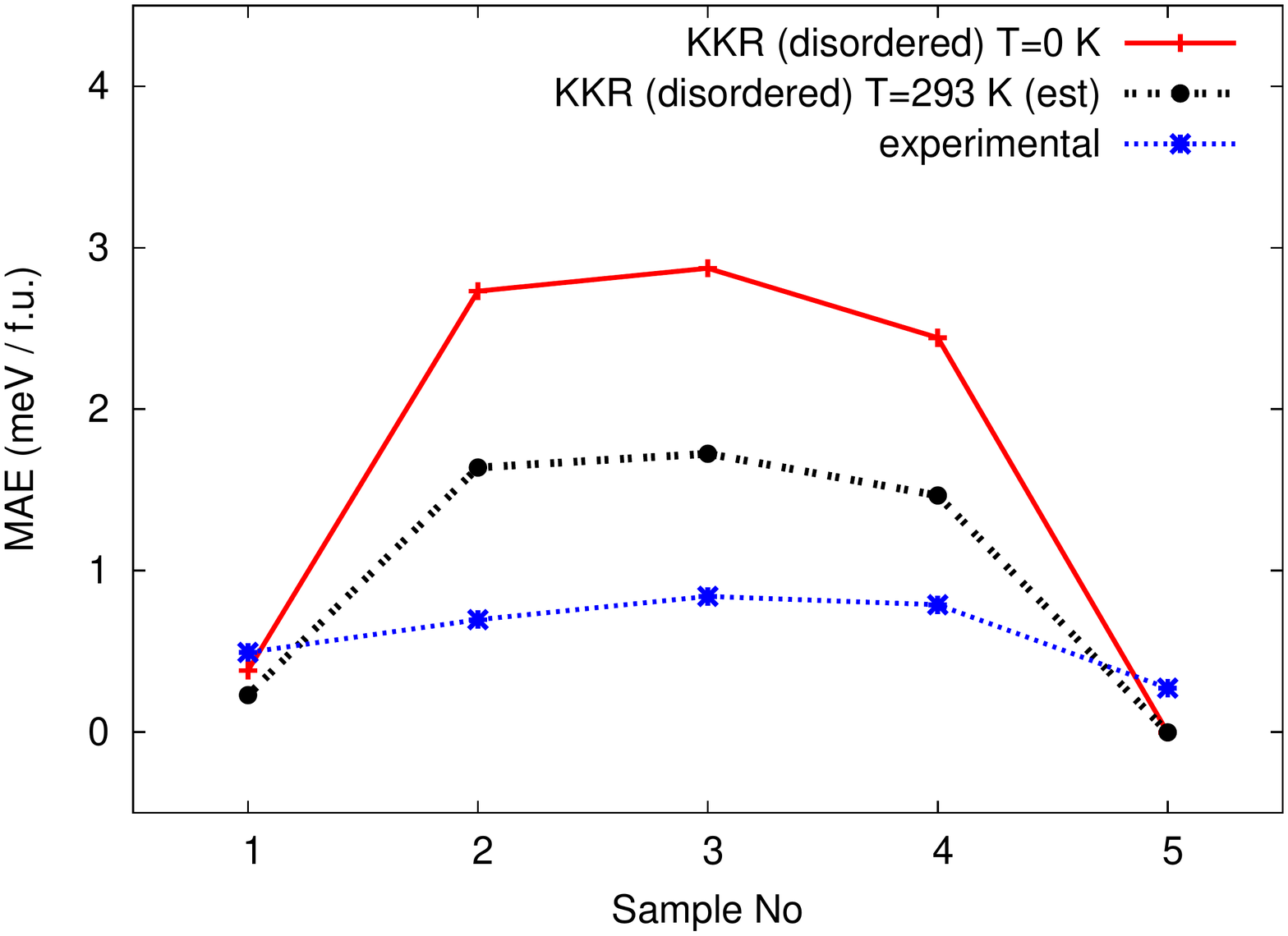}
\end{center}
\vskip -8pt
\caption{Crosses (solid line): Calculated MAE per formula unit for each of the FePt samples
in Table \ref{table:samples} modelled as partially disordered alloys with the degree of disorder
given by the experiment. Circles (dashed line):
The same values scaled down by a factor of 0.6 in order to account for temperature induced effects.
The experimental values are also displayed by stars (dotted line). \label{MAE_disordered}}
\end{figure}

Subsequently, the chemical disorder of each sample as given in Table \ref{table:samples} was taken into account using the coherent potential approximation. The corresponding results are shown in Fig.~\ref{MAE_disordered}.
In accordance with earlier work\cite{disorder2}, long-range chemical disorder drastically reduced the MAE; for sample no.~1 ($s_e=0.709$) we obtained a value of 0.4 meV/f.u., while for sample no.~5 ($s_e=0.615$) the MAE almost vanished. In fact, reducing $s$ to 0.5 can even cause a change of sign of the MAE.  In contrast, for samples no.~2 and 4 with a high degree of chemical order the MAE was reduced by less than 10 \%, and for sample no.~3  ($s_e=1$) the MAE remained unchanged with respect to our previous calculations. Taking into account again a reduction by a factor of 0.6 due to temperature effects, it is obvious that the inclusion of chemical disorder has significantly improved the agreement between experiment and theory: the trend of the MAE between the different samples is now correct and the magnitudes of the MAE are closer to the range reported by the experiment.

As mentioned above, the chemical order parameters in Table \ref{table:samples} were derived from measured diffraction intensity ratios\cite{sqrt-l10,disorder1}. However, due to an incomplete rocking curve\cite{expl}, the measured diffraction intensities, and thereby the experimentally obtained chemical disorder parameters, can only be considered approximate values. Furthermore, we note the assumption that the sample with highest MAE, sample no.~3, refers to
perfect chemical order, $s_e=1$. This seems a reasonable working hypothesis, but one worth investigating theoretically since it is central to the interpretation.

%\bigskip
\begin{figure}[htb]
\begin{center}
\includegraphics[scale=0.35, trim = 24mm 15mm 25mm 23mm, clip]{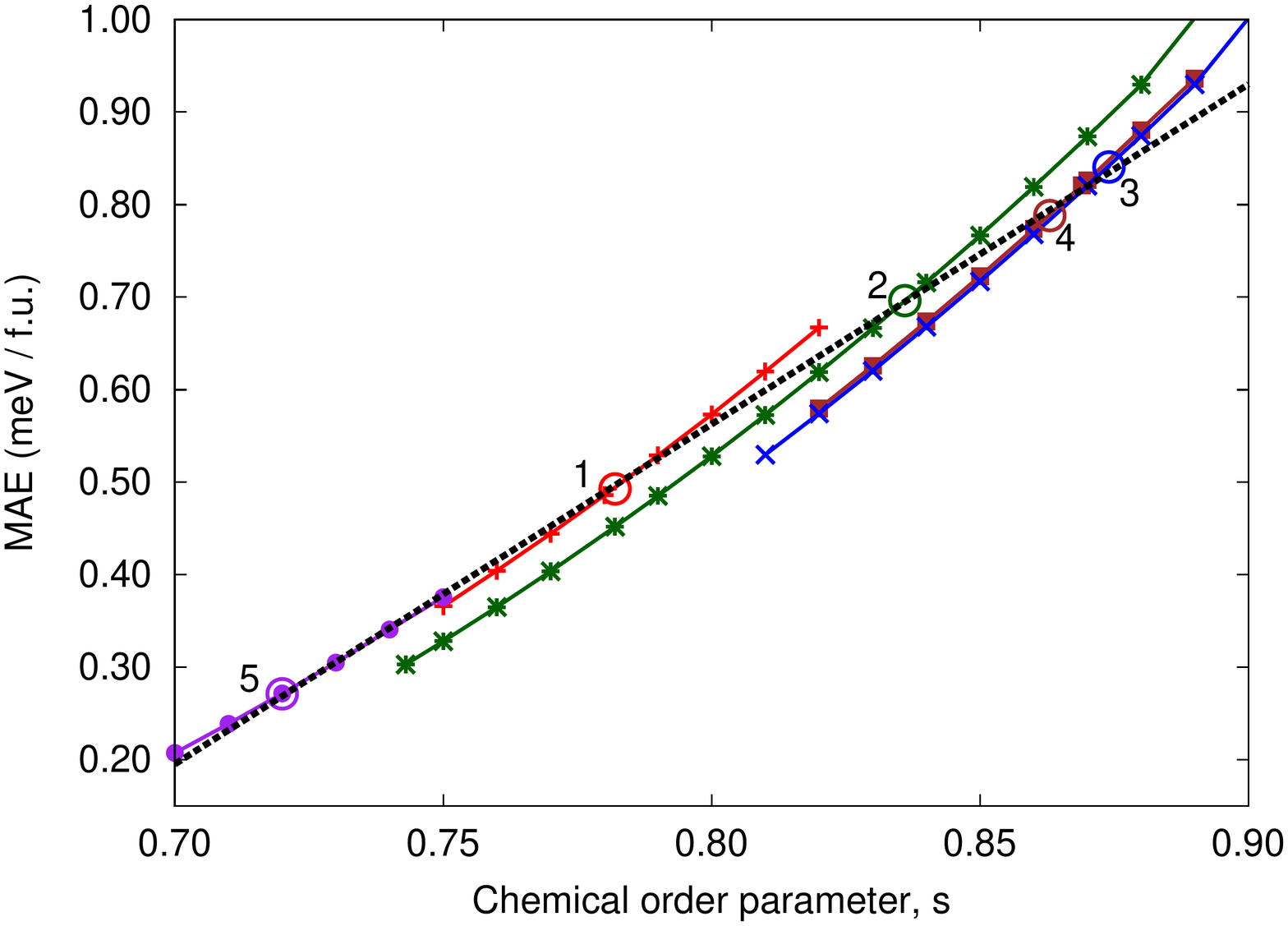}
\end{center}
\vskip -8pt
\caption{Magnetic anisotropy energy (MAE) calculated as a function of
chemical order parameter, $s$, for the FePt samples:
1 $+$, 2 $\ast$, 3  {\tiny $\blacksquare$}, 4 $\times$, 5 {\large $\bullet$}.
Solid lines serve as a guide for the eye.
Open circles are placed at the best-fit
chemical order parameter for each of the samples.  Dashed line: Linear fit.
\label{MAE_order}}
\end{figure}

The above uncertainties motivated us to perform a third set of calculations, in which the theoretical MAE was fitted to the experimental MAE using the chemical order parameter, $s$, as a fitting parameter. In Fig.~\ref{MAE_order}, for each of the samples we present the calculated MAE for an appropriate set of chemical order parameters. Firstly, for a given sample, i.e.~for fixed lattice parameters, the theoretical MAE shows a non-linear dependence on $s$. 
In Fig.~\ref{MAE_order} the circles indicate the intersection of the calculations with the experimental values for each sample as indicated. This determines the best-fit order parameter that corresponds to the experimental MAE value. As can be clearly inferred from Fig.~\ref{MAE_disordered}, for samples no.~2, 3 and 4, a smaller degree of chemical order was fitted than predicted by the experiment, namely, $s \simeq 0.836, 0.874$ and 0.863, respectively. In contrast, for samples no.~1 and 5 an increased degree of chemical order, $s \simeq  0.782$ and 0.720, was obtained. Although for a given sample the theoretical MAE shows a non-linear dependence on $s$, there is a nearly perfect
linear correlation between the experimental MAE and the best-fit chemical order parameters as indicated by the dashed line in Fig.~\ref{MAE_order}. Obviously, this remarkable linear behavior is the result of a subtle interplay of the dependence of the MAE on the lattice distortion and the chemical disorder. This is probably specific to the data set investigated here rather than being a general property.

In summary, our first principles calculations imply that lattice distortion in the FePt samples has only a minor effect on the MAE, even opposite to the experimental trend. Calculating the MAE using the highly approximate experimental chemical order parameters significantly improves the agreement between theory and experiment, in particular with regards to  the relative differences in the MAE between the samples. This indicates that the substrate-sample lattice mismatch effect on the MAE reported by Ding et al.\cite{expl} is mainly due to the variation in chemical disorder. To circumvent the uncertainty of the experimental determination of chemical disorder, we, furthermore, determined theoretical chemical order parameters that reproduced the experimental MAE values. Interestingly, a linear correlation between the MAE and the best-fit chemical order parameters is found. It should be mentioned that work is underway to perform constrained Monte-Carlo simulations of $K(T)$ for chemically disordered FePt, since this is clearly an important factor in relation to experimental data.

Financial support was provided by the Hungarian Research Foundation (contract no. OTKA K77771) and by the New Sz\'echenyi Plan of Hungary (Project ID: T\'AMOP-4.2.1/B-09/1/KMR-2010-0002). CJA  is grateful to EPSRC for the provision of a research studentship.

\end{document}